\documentclass[prd,aps]{revtex4}

\usepackage{epsfig}

\newcommand{\beq}{\begin{equation}}
\newcommand{\eeq}{\end{equation}}
\newcommand{\be}{\begin{eqnarray}}
\newcommand{\ee}{\end{eqnarray}}

\begin{document}


\preprint{
\begin{tabular}{l}
\hbox to\hsize{\hfill KAIST-TH 2005/20}\\
\hbox to\hsize{\hfill February 2006}
\end{tabular}
}

\title{
Probing CP Violating Two Higgs Doublet Model \\
through Interplay between LHC and ILC

}

\author{
 \vspace{2ex}
C. S. Kim\footnote{cskim@yonsei.ac.kr},
Chaehyun Yu\footnote{chyu@korea.ac.kr}
}

\affiliation{
\vspace{1ex}
Department of Physics, Yonsei University, Seoul 120-749, Korea
}

\author{
 \vspace{2ex}
Kang Young Lee\footnote{kylee@muon.kaist.ac.kr}
}

\affiliation{
\vspace{1ex}
Department of Physics, KAIST, Daejeon 305-701, Korea
}

\date{\today
\\}

\begin{abstract}

\noindent We study the neutral Higgs boson production
at the CERN Large Hadron Collider (LHC) and
the future $e^- e^+$ linear collider (ILC)
in the two Higgs doublet model with CP violation.
The CP-even and CP-odd scalars are mixed in this model,
which affects the production processes of neutral Higgs boson.
We examine the correlation of the Higgs production
at LHC and ILC and provide a strategy to distinguish
the model from the CP conserving model and to determine
the parameters of the Higgs sector.

\end{abstract}

\pacs{PACS numbers: 12.60.Fr,13.66.Fg,14.80.Cp}

\maketitle





\section{Introduction}

The Higgs boson is the only unobserved ingredient
of the standard model (SM),
which is responsible for the electroweak symmetry breaking (EWSB)
and the generation of fermion masses.
It is the principal motivation of the proposed future collider experiments
to search for the Higgs boson and examine
the underlying mechanism of the EWSB.
The SM Higgs boson is expected to be discovered
and its mass and production cross section to be measured
at the CERN Large Hadron Collider (LHC) \cite{lhc,djouadi},
while the complementary study on the detailed structure
of the Higgs sector will be performed
at the future $e^- e^+$ linear collider (ILC) \cite{djouadi}.

Conventional wisdom is that the standard model
is not the final theory but just an effective theory
of the fundamental structure.
The Higgs sector is generically extended when we consider
the new physics beyond the SM because the new physics
usually has larger symmetry than the SM and then more
symmetry breaking is required than the minimal EWSB.
The two Higgs-doublet (2HD) model is one of the simplest
extension of the Higgs sector, which consists of two scalar SU(2) doublets.
In the 2HD model, the physical states are three neutral scalars
and a pair of charged scalars after the Goldstone modes are eaten up by
the $W$ and $Z$ bosons.
When both Higgs doublets couple to all fermions,
their Yukawa coupling matrices cannot be simultaneously diagonalized
with the mass matrices of fermions.
Then the off-diagonal matrix elements exist in general,
which gives rise to the flavor-changing neutral current (FCNC)
mediated by a neutral Higgs boson at tree level.
Thus one has to constrain the Yukawa couplings arbitrarily,
or introduce another symmetry to suppress the FCNC,
$e.g.$ an approximate flavor symmetry \cite{afs}.
The phenomenological implications of the FCNC to the collider signal
and flavor physics have been widely studied
\cite{collider,flavour}.
A natural flavor conservation (NFC) has been suggested
in order to avoid the dangerous FCNC
by imposing a discrete symmetry on the Higgs and fermion fields \cite{GW}.
We choose the discrete symmetry such that one Higgs doublet
couples to up-type quarks
and the other doublet couples to down-type quarks
to get rid of the tree level FCNC.
This is also the Higgs structure of the minimal supersymmetric
standard model (MSSM).
Such a discrete symmetry on the Higgs doublet fields forbids
the Higgs sector to contain the CP violating terms.
Therefore, the CP is the manifest symmetry of the theory and
the physical states of neutral scalars
are two CP-even Higgs bosons $H$ and $h$ and one CP-odd Higgs boson $A$.

Assuming the breakdown of the discrete symmetry in the Higgs sector,
complex Higgs self-couplings exist in general, and consequently
the explicit and/or spontaneous CP violation is allowed in the Higgs sector.
In such a case $H$, $h$ and $A$ are no more CP eigenstates
and the CP-even and CP-odd states are mixed.
The Higgs boson couplings to gauge bosons and fermions also
depend upon the mixing angle between the CP-even and CP-odd
Higgs bosons and so does the production of neutral Higgs bosons
at the future colliders.
Even it is possible that the production cross section
of the Higgs boson is suppressed to be missed at colliders
in some parameter space
although the Higgs boson mass is low enough to be produced at such colliders.
However, if one neutral Higgs boson production may be suppressed,
other Higgs boson production processes will be enhanced instead
according to sum rules of Higgs couplings
and we can find the evidence of Higgs boson in general.
Analysis to search for the neutral Higgs boson at the $e^- e^+$ collider
in the CP violating 2HD model
has been performed in the Refs. \cite{gunion,ginzburg,kim}.
In this work, we consider the Higgs sector with soft violation
of the discrete symmetry in the 2HD model and
discuss its implication for the neutral Higgs boson production
at LHC and ILC.

This paper is organized as follows: In section 2,
we briefly review the general 2HD model with non-zero CP violation
and define the physical neutral Higgs bosons.
Possible scenarios are examined in section 3
and correlation of LHC and ILC is discussed in section 4.
Finally we conclude in section 5.

\section{The Model}

The general Higgs potential of the 2HD model is given by
\be
V &=& \frac{1}{2} \lambda_1 (\phi_1^\dagger \phi_1)^2
   + \frac{1}{2} \lambda_2 (\phi_2^\dagger \phi_2)^2
   + \lambda_3 (\phi_1^\dagger \phi_1)(\phi_2^\dagger \phi_2)
   + \lambda_4 (\phi_1^\dagger \phi_2)(\phi_2^\dagger \phi_1)
\nonumber
\\
&& + \frac{1}{2} [ \lambda_5 (\phi_1^\dagger \phi_2)^2 + {\rm H.c.}]
   + [ \lambda_6 (\phi_1^\dagger \phi_1)(\phi_1^\dagger \phi_2)
     + \lambda_7 (\phi_2^\dagger \phi_2)(\phi_1^\dagger \phi_2)
     + {\rm H.c.}]
\nonumber
\\
&& - m_{11}^2 (\phi_1^\dagger \phi_1)
   - m_{22}^2 (\phi_2^\dagger \phi_2)
   - [ m_{12}^2 (\phi_1^\dagger \phi_2) + {\rm H.c.} ],
\ee
where $\lambda_5, \lambda_6, \lambda_7$ and $m_{12}^2$ are
complex parameters and others are real.
The discrete symmetry $\phi_1 \to -\phi_1$ or $\phi_2 \to -\phi_2$ is
imposed in order to avoid a dangerous FCNC,
which leads to the absence of $m_{12}^2$, $\lambda_6$ and $\lambda_7$.
We allow soft violation of the discrete symmetry by the dimension 2
terms $m_{12}^2 \ne 0$ in this work for the explicit CP violation.

The potential is minimized with the vacuum expectation values
\be
\langle \phi_1 \rangle = \frac{1}{\sqrt{2}}
\left( \begin{array}{c}
             0 \\
             v_1
           \end{array}   \right),~~~~
\langle \phi_2 \rangle = \frac{1}{\sqrt{2}}
\left( \begin{array}{c}
             0 \\
             v_2 e^{i \xi}
           \end{array}   \right),
\ee
where $v_1^2 + v_2^2 = v^2 = 4 m_W^2/g^2$ and
the phase $\xi$ leads to the spontaneous CP violation.
The minimization of the potential (1) at $\langle \phi_1 \rangle$
and $\langle \phi_2 \rangle$ yields the relation
\be
{\rm Im} ( m_{12}^2 e^{i \xi})
     = v_1 v_2 {\rm Im} ( \lambda_5 e^{2 i \xi} ).
\ee
The global transform $\phi_i \to \phi_i e^{i \varphi_i}$
preserves the potential invariant with the rephasing;
$\lambda_5 \to \lambda_5 e^{-2i(\varphi_2-\varphi_1)}$,
 $m_{12}^2 \to m_{12}^2 e^{-i(\varphi_2-\varphi_1)}$
and $\xi \to \xi + \varphi_2 - \varphi_1$
while $\lambda_i~(i=1,\cdot \cdot \cdot ,4)$ and $m_{11,22}^2$
are not changed.
Thus we can choose $\xi=0$ indicating no spontaneous CP violation
but the wholly explicit CP violation.
Then the parameter ${\rm Im} ~m_{12}^2$ can be replaced by
$ {\rm Im} ~ \lambda_5$, which plays
the role of the order parameter of the CP violation.

The charged states are the Goldstone mode
$G^\pm = \phi_1^\pm \cos \beta + \phi_2^\pm \sin \beta$,
and the charged Higgs mode
$H^\pm = -\phi_1^\pm \sin \beta + \phi_2^\pm \cos \beta$
with the mass $m^2_{H^\pm}= [{\rm Re}~m_{12}^2 /v_1 v_2
-\lambda_4 - {\rm Re}~\lambda_5] v^2/2$
and $\tan \beta = v_2/v_1$ .
In this paper, we concentrate on the neutral Higgs boson production
and do not pay attention to the charged states.
In order to pick out the physically allowed set of the Higgs potential
parameters, however, we have to assume the charged Higgs boson mass
in our numerical analysis.

The neutral states are defined by
\be
G^0 &=& \sqrt{2}
    ( {\rm Im}~ \phi_1^0 \cos \beta +  {\rm Im}~ \phi_2^0 \sin \beta ),
\nonumber \\
A^0 &=& \sqrt{2}
    ( -{\rm Im}~ \phi_1^0 \sin \beta +  {\rm Im}~ \phi_2^0 \cos \beta ),
\nonumber \\
\varphi_1 &=& \sqrt{2} {\rm Re}~ \phi_1^0,
\nonumber \\
\varphi_2 &=& \sqrt{2} {\rm Re}~ \phi_2^0.
\ee
The mass matrix of neutral Higgs bosons is constructed as the form
\be
{\cal M}^2 =
v^2
\left( \begin{array}{ccc}
R \sin^2 \beta + \lambda_1 \cos^2 \beta &
\left(\lambda_3 + \lambda_4 + {\rm Re} ~\lambda_5 - R \right)
     \frac{v_1 v_2}{v^2} &
-\frac{1}{2} {\rm Im} ~\lambda_5 v^2 \sin \beta
\\
\left(\lambda_3 + \lambda_4 + {\rm Re} ~\lambda_5 - R \right)
     \frac{v_1 v_2}{v^2} &
R \cos^2 \beta + \lambda_2 \sin^2 \beta &
-\frac{1}{2} {\rm Im} ~\lambda_5 v^2 \cos \beta
\\
-\frac{1}{2} {\rm Im} ~\lambda_5 v^2 \sin \beta &
-\frac{1}{2} {\rm Im} ~\lambda_5 v^2 \cos \beta &
R - {\rm Re} ~\lambda_5
           \end{array}   \right),
\ee
where $ R = {\rm Re}~m_{12}^2 /v_1 v_2$.
The non-zero off-diagonal elements ${\cal M}^2_{13}$ and ${\cal M}^2_{23}$
indicate the CP violation, since three neutral scalars are mixed
and no more CP eigenstates.
The parameter $ {\rm Im} ~\lambda_5$ is the only CP violating parameter
in this model.

We diagonalize the mass matrix by the orthogonal transformation
\be
{\cal M}^2_d = {\cal R} {\cal M}^2 {\cal R}^\dagger,
\ee
where the orthogonal matrix ${\cal R}$ can be parameterized
by 3 Euler angles $\theta_a$,  $\theta_b$, $\theta_c$
\be
{\cal R} &=&
\left( \begin{array}{ccc}
1 & 0 &0 \\
0 &c_c &s_c \\
0 & -s_c& c_c \\
           \end{array}   \right)
\left( \begin{array}{ccc}
c_b&0&s_b \\
0&1&0 \\
-s_b&0&c_b \\
           \end{array}   \right)
\left( \begin{array}{ccc}
-s_a&c_a&0 \\
c_a&s_a&0 \\
0&0&1 \\
           \end{array}   \right)
\nonumber \\
&=&
\left( \begin{array}{ccc}
-c_b s_a&c_a c_b&s_b \\
c_a c_c+s_a s_b s_c&s_a c_c-c_a s_b s_c&c_b s_c \\
-c_a s_c+ s_a s_b c_c&-s_a s_c-c_a s_b c_c&c_b c_c
           \end{array}   \right),
\ee
where $s_{a,b,c} = \sin \theta_{a,b,c}$
and $c_{a,b,c} = \cos \theta_{a,b,c}$.
The choice of the angle $\theta_a$ is different from other angles
to follow the convention of the mixing angle between
CP-even Higgs bosons.
Hereafter we set $\alpha \equiv \theta_a$ by convention.
Then the physical states for neutral Higgs bosons
$h_1, h_2, h_3$ are defined by
\be
\left( \begin{array}{c}
h_1\\
h_2\\
h_3
       \end{array}   \right)
   = {\cal R}
\left( \begin{array}{c}
\varphi_1\\
\varphi_2\\
A
       \end{array}   \right).
\ee
The CP-odd state $A$ is mixed with CP-even states $\varphi_1, \varphi_2$
and it indicates a manifest CP violation in the neutral Higgs sector.

The Yukawa couplings are given by with the discrete symmetry
\be
{\cal L}_Y = -g_{ij}^d \bar{Q}_L^i \phi_1 d_R^j
- g_{ij}^d \bar{L}_L^i \phi_1 l_R^j
- g_{ij}^u \bar{Q}_L^i \tilde{\phi}_2 u_R^j + {\rm H.c.},
\ee
where $\tilde{\phi}_2 = i \tau_2 \phi_2$.
The relevant terms for the dominant production process $p p \to g g \to h$
at LHC is the top Yukawa coupling given by
\be
{\cal L}_{Y_t} = -\frac{g}{2} \frac{m_t}{m_W }
   \sum_{i=1}^3
   \left( \frac{({\cal R}^{-1})_{2i}}{\sin \beta}
     +i \frac{({\cal R}^{-1})_{3i} \cos \beta}{\sin \beta} \right)
        \bar{t}_L t_R h_i + {\rm H.c.}.
\ee

\section{Scenarios for neutral Higgs boson production}

If ${\rm Im} ~\lambda_5=0$,
we can see that the mass matrix of Eq. (5) is reduced
to that of the CP conserving case,
where the CP-even and CP-odd Higgs bosons are separated.
This is corresponding to the case of $\theta_b = \theta_c = 0$
in the matrix ${\cal R}$.
A few interesting scenarios for nonzero but limiting values of
$\theta_b$ and $\theta_c$ are considered at each experiment
in this section.

\subsection{the CERN Large Hadron Collider (LHC)}

We expect that the SM Higgs boson will be discovered at LHC
and its mass and the cross section will be measured.
The dominant production channel of the neutral Higgs boson
at LHC is the gluon fusion process, $p p \to g g \to h$.
Since the cross section for the neutral Higgs boson
production crucially depends upon Higgs couplings,
the large deviation of the measured cross section from the SM prediction
indicates the evidence of the new structure of the Higgs sector.
Even it is possible that we cannot discover the lightest Higgs boson
due to the suppressed coupling with some combination of
$\alpha$, $\theta_b$, $\theta_c$ and $\tan \beta$
although the Higgs mass is small enough to be observed at LHC.
Here, we examine a scenario where this dominant channel
($p p \to g g \to h$) is blind at LHC for three neutral Higgs bosons.
The cross section for the lightest neutral Higgs boson is given by
\be
\sigma(g g \to h_1) = \left(
            \frac{\cos^2 \alpha \cos^2 \theta_b}{\sin^2 \beta}
              +\frac{\cos^2 \beta \sin^2 \theta_b}{\sin^2 \beta} \right)
                 \cdot  \sigma_{\rm SM}(m_{h_1}).
\ee
The orthogonality of the matrix ${\cal R}$ gives rise to a sum rule
for ratios of cross sections as
\be
\sum_{i=1}^3
\frac{\sigma(gg \to h_i)}{\sigma_{\rm SM}(m_{h_i})}
 = \frac{1 + \cos^2 \beta}{\sin^2 \beta} > 1,
\ee
where $\sigma_{\rm SM}(m)$ is the SM cross section
with the Higgs mass $m$.
The cross section $\sigma(gg \to h_1)$ vanishes
in the limit of $\cos^2 \alpha \to 0$,
$ \sin^2 \theta_b \to 0$.
In this limit, the cross section for $h_2$ is given by
\be
\sigma(g g \to h_2) = \left(
            \frac{\cos^2 \theta_c}{\sin^2 \beta}
              +\frac{\cos^2 \beta \sin^2 \theta_c}{\sin^2 \beta} \right)
                 \cdot  \sigma_{\rm SM}(m_{h_2}).
\ee
Thus if $\cos^2 \theta_c \to 0$ and $\tan \beta$ is sufficiently large,
$\sigma(g g \to h_2)$ is also suppressed and we cannot find $h_2$
through $ p p \to g g \to h$ process at LHC.
Finally we consider the heaviest Higgs boson $h_3$.
Since the first and second cross sections are suppressed,
the cross section $\sigma(g g \to h_3)$
should be larger than the SM cross section of the corresponding mass
due to the sum rule of Eq. (12),
and $h_3$ should be observed through this channel.
Now if we assume that $m_{h_3}$ is too heavy to be produced
at LHC, then no neutral Higgs boson is observed at LHC
through $ p p \to g g \to h$ process.
Consequently neutral Higgs bosons are blind
through this channel at LHC in such a  scenario.

Let us test the above `blind' scenario in more detail.
Actually $m_3$ is not a free parameter in this model
when other Higgs boson masses and mixing angles are fixed.
In the Fig. 1, we show the value of $m_3$ with respect to $\tan \beta$
in the limit considered here.
We find that $m_3$ is bounded above, and such a complete blind scenario
is not plausible.
In practice, we need not take a strict limit of
$\cos \theta / \sin \theta \to 0$,
but it is enough if the signal is too small to be extracted
from the large background of the hadron collider.
We set the condition, $\cos^2 \alpha$, $\sin^2 \theta_b$,
$\cos^2 \theta_c < 0.01$, in this analysis.
Higgs boson masses $m_1$ and $m_2$ are varied up to 1 TeV
which is the experimental bound for the SM Higgs boson
to be found at LHC.
For the numerical analysis,
we demand the following constraints on the model parameters;
(1) the perturbativity on the quartic couplings, $\lambda_i/4 \pi <1$,
(2) the ordering of Higgs masses, $m_1 < m_2 <m_3$.
If $\tan \beta$ is large, the $b$ quark Yukawa coupling is also large
so that the contribution of $b$ quark loop becomes important.
Including the contribution of $b$ quark loop contaminates
the relations in our discussion.
However, as we can see in Fig. 1, the favored value of $\tan \beta$
is less than 10, and we can safely assume that the top quark loop
dominates in $pp \to gg \to h$ process.

\subsection{the future $e^- e^+$ linear collider (ILC)}

The phenomenology of the Higgs sector at the $e^- e^+$ linear colliders
is governed by couplings of the Higgs bosons to gauge bosons.
We write generalized $h_i ZZ$ couplings here:
\be
h_1 ZZ &\propto& \sin (\beta-\alpha) \cos \theta_b,
\nonumber
\\
h_2 ZZ &\propto& \cos (\beta-\alpha) \cos \theta_c
              - \sin (\beta-\alpha) \sin \theta_b \sin \theta_c,
\nonumber
\\
h_3 ZZ &\propto& - \cos(\beta-\alpha) \sin \theta_c
              - \sin (\beta-\alpha) \sin \theta_b \cos \theta_c,
\ee
which are normalized by the SM coupling $g m_Z/\cos \theta_W$.
Due to the orthogonality of ${\cal R}$,
these couplings satisfy a few sum rules,
which can be found in the literatures \cite{gunion,pomarol}.

If we assume that $\sin \theta_b \sim 0$ and $\cos \theta_c \sim 0$,
$h_2$ is decoupled and identified with the CP-odd Higgs boson $A$.
In the limit that $\cos \theta_b \sim \cos \theta_c \sim 0$,
$h_1$ is decoupled to be $A$.
In both cases, Higgs-gauge couplings $g_{_{h_i ZZ}}$
and $g_{_{h_i h_j Z}}$ go close to those of the CP conserving case.
Thus these limiting cases are similar to the CP conserving case
except for the possibility that the lightest Higgs may be
the CP-odd Higgs boson.

More interesting scenario is obtained by taking the limit
$\sin \theta_c \to 0$.
In this case, the off-diagonal elements of the mass matrix
of neutral Higgs bosons become
\be
{\cal M}^2_{13} &=& s_a c_b s_b (m_3^2-m_2^2),
\nonumber \\
{\cal M}^2_{23} &=& -c_a c_b s_b (m_3^2-m_2^2).
\ee
Comparing the ratio ${\cal M}^2_{13}/{\cal M}^2_{23}$
from Eq. (15) with the same ratio given in Eq. (5)
leads to $\tan \beta = - \tan \alpha$ and thus $\beta = - \alpha$.
The CP violating parameter ${\rm Im}~\lambda_5$
is directly related to $\theta_b$ and Higgs masses,
\be
{\rm Im}~\lambda_5 = \sin 2 \theta_b \frac{m_3^2-m_2^2}{v^2}.
\ee
If we additionally assume that $\sin \theta_b$ is close to 1,
the lightest Higgs boson decouples to be the CP-odd Higgs
and heavier Higgses $h_2$ and $h_3$ are mixed with each other.
The $g_{_{h_i ZZ}}$ and $g_{_{h_i h_j Z}}$ couplings are given by
\be
h_1 ZZ &\propto& \epsilon_b \sin (\beta-\alpha),
\nonumber
\\
h_2 ZZ &\propto& \cos (\beta-\alpha),
\nonumber
\\
h_3 ZZ &\propto& - \sin ( \beta-\alpha),
\ee
where $\epsilon_b = \cos \theta_b$.
This limit may look like the CP conserving case
since the CP-odd Higgs decouples.
However, we see that the ratio
$| g_{_{h_2 ZZ}}/g_{_{h_3 ZZ}} | = 1/ \tan (\beta-\alpha)$,
while $| g_{_{h ZZ}}/g_{_{H ZZ}}|  = \tan (\beta-\alpha)$
for the CP conserving case.
It may be a clue to discriminate this scenario
from the CP conserving model in the gauge-Higgs sector
without the manifest observation of the CP asymmetry \cite{kim}.


\section{Interplay between LHC and ILC}

It is natural that the first evidence of the Higgs boson
will be discovered at LHC before ILC.
Thus we can assume that the lightest Higgs boson mass and
corresponding cross section will be measured at LHC
without loss of generality.
We define $\Delta$ to be the ratio of the cross section of
$p p \to g g \to h$ in our model to the SM cross section
of corresponding $m_h$,
\be
\Delta \equiv \frac{\sigma(p p \to g g \to h_1)}
                   {\sigma^{pp}_{\rm SM}(m_{h_1})},
\ee
where $\sigma^{pp}_{\rm SM}(m_{h_1})$ is the SM cross section of
$p p \to g g \to h$ channel with the Higgs boson mass $m_{h_1}$.
The mixing angle $\theta_b$ is expressed in terms of
$\Delta$ and other angles such as
\be
\cos^2 \theta_b = \frac{\Delta \sin^2 \beta - \cos^2 \beta}
                       {\cos^2 \alpha - \cos^2 \beta},
\ee
when $\cos^2 \alpha \ne \cos^2 \beta$.
If the measured cross section deviates from the SM prediction,
$\Delta \ne 1$, it directly indicates a new structure of the Higgs sector.

The most promising channel at  ILC is
the Higgsstrahlung process $e^- e^+ \to Z h_i$.
In the CP conserving model, the CP-odd Higgs boson $A$
does not couple to the gauge boson.
So the observation of $Z A$ production is a direct evidence
of the CP violating model in principle.
However, it is hard to tag $A$ or $h$ in the Higgsstrahlung process
due to its spin 0 nature.
Moreover in our case, the produced neutral Higgs boson is not
a pure CP eigenstates but a mixed one.
Thus it is impossible to find the $Z A$ production in our model.
Let us first assume that we have the minimal information of the Higgs sector
from LHC at the moment when the $e^- e^+$ linear collider starts running:
$e.g.$ the lightest Higgs mass $m_{h_1}$ and the ratio of the cross section
$\sigma(p p \to g g \to h_1)/\sigma^{pp}_{\rm SM}$
have been already determined.
The cross section of $e^- e^+ \to Z h_i$ is
\be
\sigma (e^+ e^- \to h_i Z) = f_i^2 ~ \sigma^e_{\rm SM} (m_{h_i})
\ee
where $f_{i}$ are the normalized $h_i ZZ$ couplings given in Eq. (14)
and $\sigma^e_{\rm SM}(m_{h_i})$ is the SM cross section for
$e^+ e^- \to h_i Z$ process with corresponding Higgs mass.
For given $m_{h_1}$ and $\Delta$, we can rewrite
\be
f_1^2 = \sin^2 (\beta - \alpha) \frac{ \Delta \sin^2 \beta - \cos^2 \beta}
                     {\cos^2 \alpha - \cos^2 \beta} ,
\ee
assuming $\cos^2 \alpha \ne \cos^2 \beta$.
In Fig. 2, we plot the cross section $\sigma(e^- e^+ \to Z h_1)$
with respect to the measured $\Delta$.
Values of $\tan \beta$ is assumed to be fixed since $\tan \beta$
can be measured in separate processes, $e.g.$ charged Higgs sector,
while $\alpha$ is varied.

Next we assume that we have already measured cross sections
for 2 channels at LHC,
$\sigma(g g \to h_1)$ and $\sigma(g g \to h_2)$,
and, however, measured only one cross section at  ILC.
This is a very likely assumption since the CM energy of  ILC
is much lower than LHC.
In such a scenario, we can critically discriminate the CP violating model
with CP conserving model.
In the CP conserving model, $\theta_b$, $\theta_c =0$,
the cross sections at LHC are
\be
\Delta_{\rm LHC1} \equiv \frac{\sigma(g g \to h_1)}{\sigma_{\rm SM}(m_{h_1})}
         = \frac{\cos^2 \alpha}{\sin^2 \beta},~~~~~~~
\Delta_{\rm LHC2} \equiv \frac{\sigma(g g \to h_2)}{\sigma_{\rm SM}(m_{h_2})}
         = \frac{\sin^2 \alpha}{\sin^2 \beta},
\ee
while the cross section at the ILC is given by
\be
\Delta_{\rm ILC} \equiv \frac{\sigma(e^- e^+ \to h_1 Z)}
                             {\sigma_{\rm SM}(m_{h_1})}
                = \sin^2 (\beta - \alpha).
\ee
Then these observables satisfy the relation
\be
\Delta_{\rm LHC2} (1 - \Delta_{\rm ILC})
    = ( \sqrt{\Delta_{\rm LHC1} \Delta_{\rm ILC}} -1)^2,
\ee
which is the surface on $(\Delta_{\rm LHC1}, \Delta_{\rm LHC2},
\Delta_{\rm ILC})$ plane.
Since this relation is derived in the CP conserving two Higgs doublet model,
a deviation of any observable from this relation indicates
the CP violating model and/or more complex Higgs structure.
In Fig. 3, we show a contour plot on this relation.
We assume that the measurement of $\Delta_{\rm LHC1}$
agrees with the SM prediction.
The contour is corresponding to $\theta_b = \theta_c = 0$
and we plot the departure with
$\theta_b \ne 0$ and $\theta_c = 0$ case.
The $\theta_c \ne 0$ and $\theta_b = 0$ provides
a line perpendicular to this plane.


\section{Concluding Remarks}

The neutral Higgs boson production processes
have been explored together at LHC and ILC
in the context of the two Higgs-doublet model with non-zero CP violation.
Because of the spin 0 nature, it is hard to get
the direct CP violating signals from the neutral Higgs sector.
Instead, by investigating the combined production rate at LHC and ILC,
we can find out if the CP violating signal to indicate
the scalar-pseudoscalar mixing.
We suggest a few limiting cases which shows characteristic phenomenology
in the neutral Higgs boson production and a strategy to
determine the Higgs mixing angles at  ILC with the data obtained
from LHC.
\\

\acknowledgments
The work of C.S.K. was supported
in part by  CHEP-SRC Program and
in part by the Korea Research Foundation Grant funded by
the Korean Government (MOEHRD) No. KRF-2005-070-C00030.
The work of K.Y.L. was supported by Korea Research Foundation Grant
(KRF-2003-050-C00003).

\newpage

\def\PRD #1 #2 #3 {Phys. Rev. D {\bf#1},\ #2 (#3)}
\def\PRL #1 #2 #3 {Phys. Rev. Lett. {\bf#1},\ #2 (#3)}
\def\PLB #1 #2 #3 {Phys. Lett. B {\bf#1},\ #2 (#3)}
\def\NPB #1 #2 #3 {Nucl. Phys. {\bf B#1},\ #2 (#3)}
\def\ZPC #1 #2 #3 {Z. Phys. C {\bf#1},\ #2 (#3)}
\def\EPJ #1 #2 #3 {Euro. Phys. J. C {\bf#1},\ #2 (#3)}
\def\JHEP #1 #2 #3 {JHEP {\bf#1},\ #2 (#3)}
\def\IJMP #1 #2 #3 {Int. J. Mod. Phys. A {\bf#1},\ #2 (#3)}
\def\MPL #1 #2 #3 {Mod. Phys. Lett. A {\bf#1},\ #2 (#3)}
\def\PTP #1 #2 #3 {Prog. Theor. Phys. {\bf#1},\ #2 (#3)}
\def\PR #1 #2 #3 {Phys. Rep. {\bf#1},\ #2 (#3)}
\def\RMP #1 #2 #3 {Rev. Mod. Phys. {\bf#1},\ #2 (#3)}
\def\PRold #1 #2 #3 {Phys. Rev. {\bf#1},\ #2 (#3)}
\def\IBID #1 #2 #3 {{\it ibid.} {\bf#1},\ #2 (#3)}

\begin{center}
\begin{figure}[b]
\vskip 3cm
\hbox to\textwidth{\hss\epsfig{file=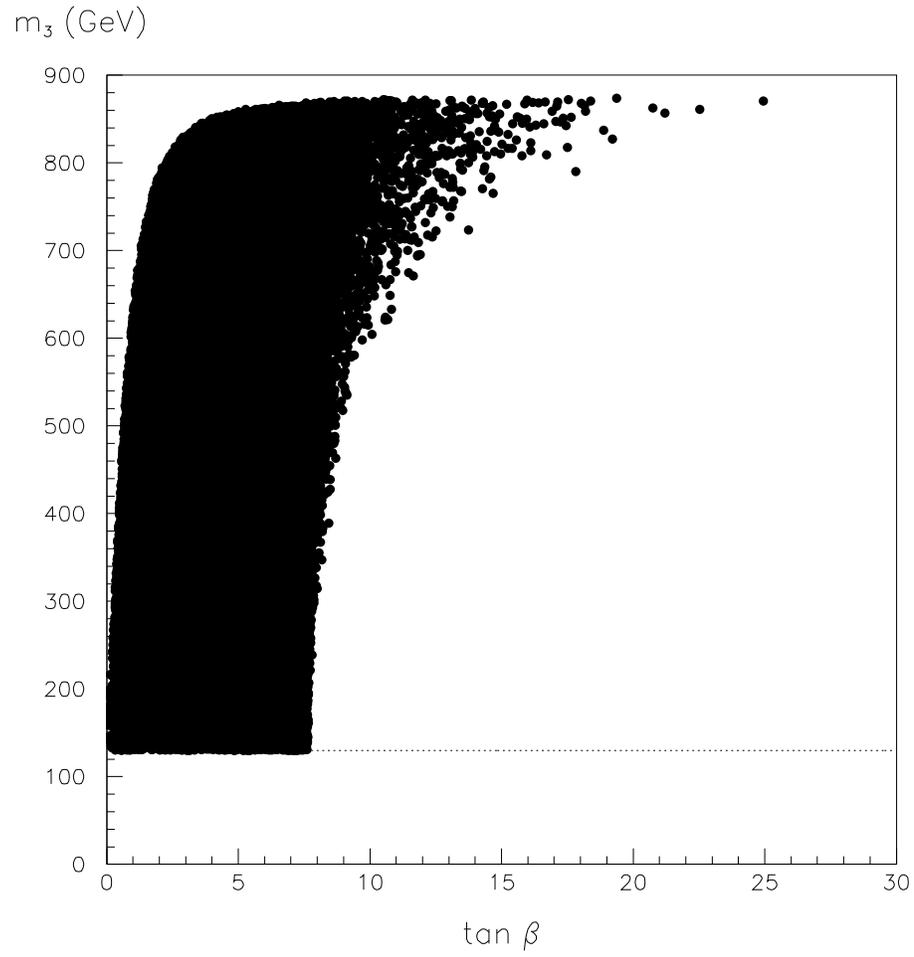,width=14cm}\hss}
\vspace{1cm}
\caption{
In the blind scenario at LHC, the allowed values of $m_{h_3}$
with respect to $\tan \beta$.
}
\end{figure}
\end{center}

\newpage

$~$
\vskip 3cm

\begin{center}
\begin{figure}[b]
\hbox to\textwidth{\hss\epsfig{file=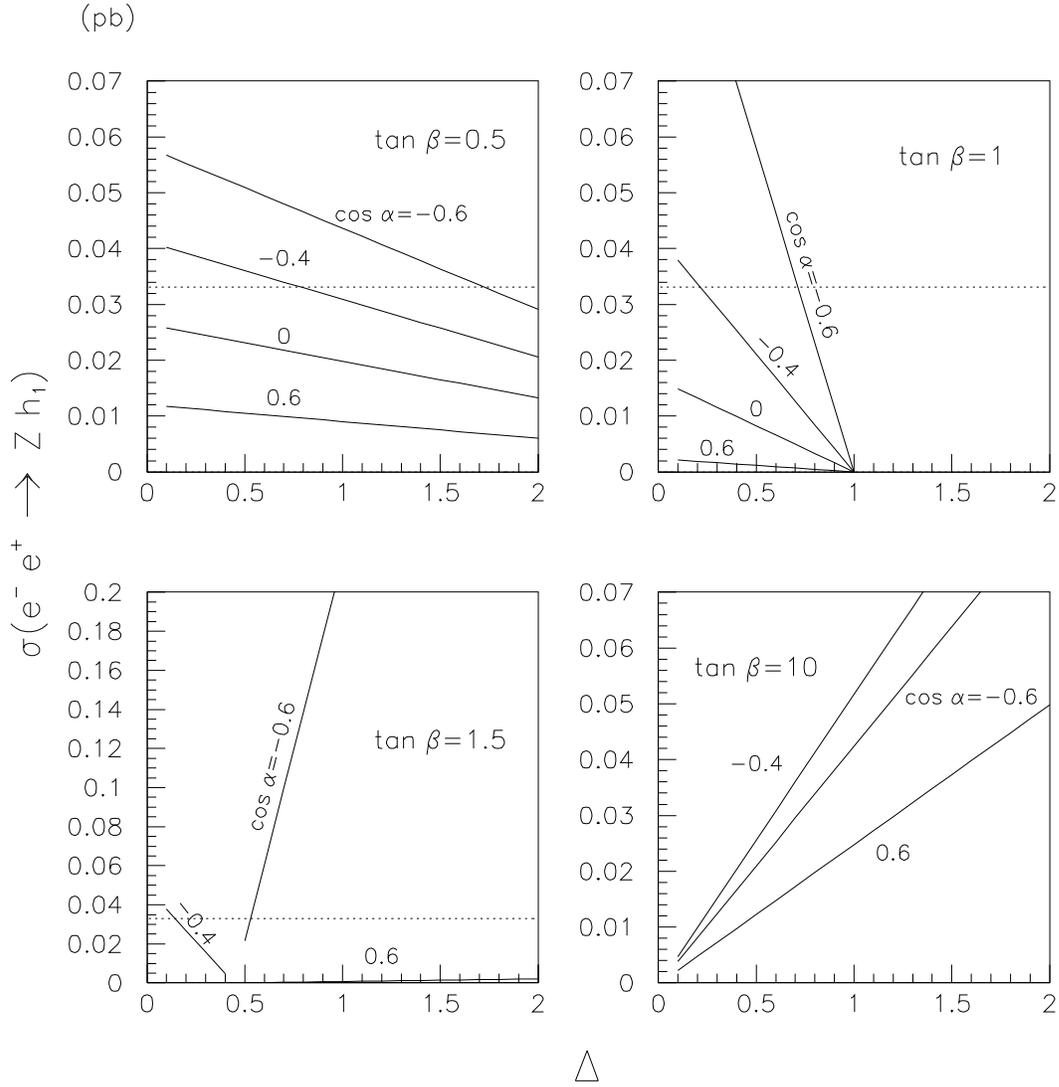,width=16cm}\hss}
\vspace{1cm}
\caption{
Predicted cross section of $e^- e^+ \to Z h_1$ process
with respect to $\Delta$ obtained from LHC.
$\tan \beta$ and  the mixing angle $\alpha$ is varied.
The Higgs mass is assumed to be 130 GeV and the dotted
horizontal line is corresponding cross section predicted by the SM.
}
\end{figure}
\end{center}

\newpage

\begin{center}
\begin{figure}[b]
\hbox to\textwidth{\hss\epsfig{file=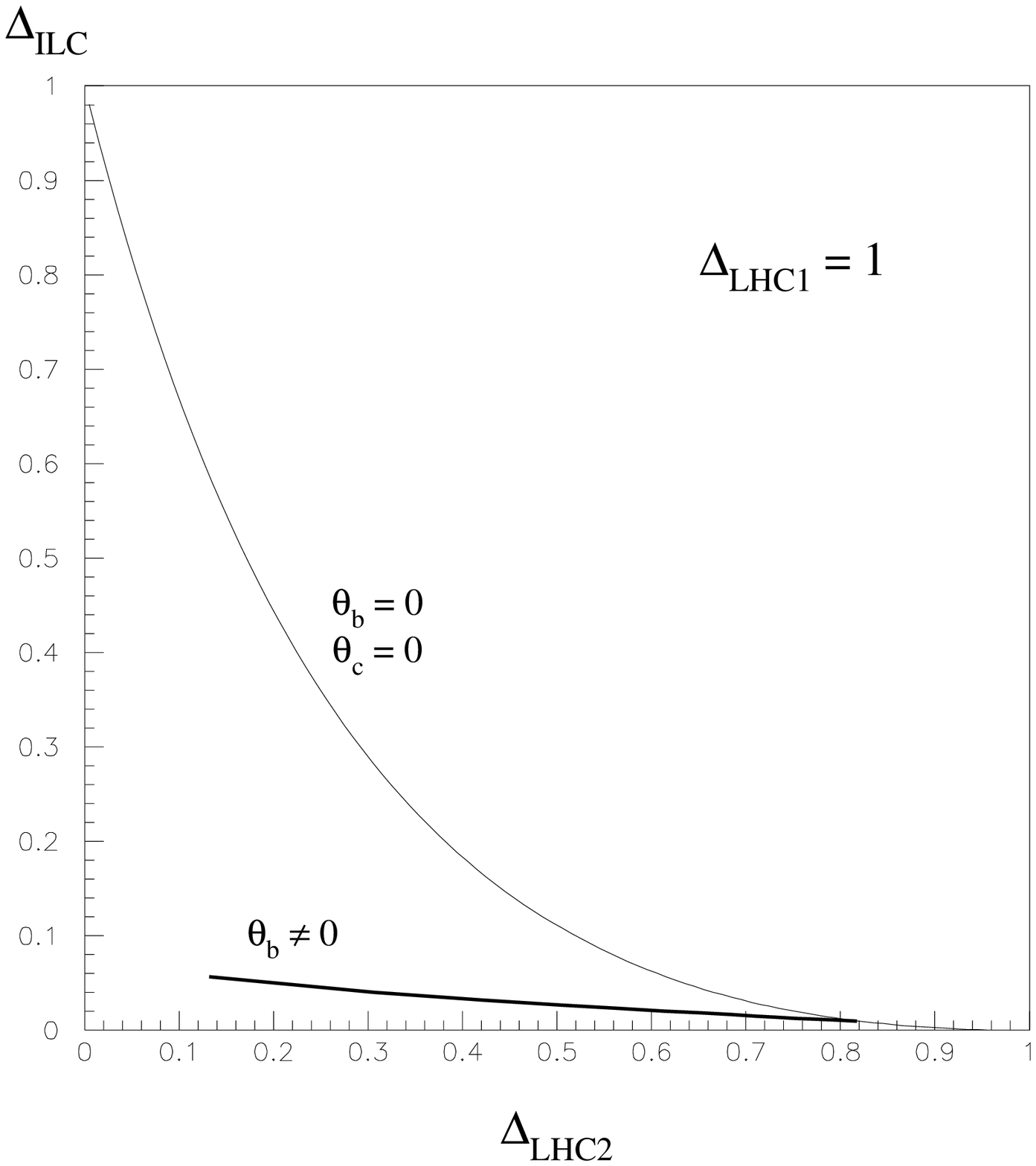,width=17cm}\hss}
\vspace{1cm}
\caption{
Deviation from the CP conserving limit
in $(\Delta_{\rm LHC 2},\Delta_{\rm ILC})$ plane.
$\Delta_{\rm LHC 1} = 1$ is assumed.
}
\end{figure}
\end{center}


\begin{thebibliography}{99}

\bibitem{lhc} ATLAS Collaboration,
{\sl ATLAS : Detector and Physics performance Technical Design Report}
Vol. II, Report No. CERN/LHCC 99-15 (1999).

\bibitem{djouadi} {\it See} A. Djouadi, [arXiv] hep-ph/0503172;
[arXiv] hep-ph/0503173
and references therein.

\bibitem{afs}
L. J. Hall and S. Weinberg, \PRD 48 979 1993 ;
Y.-L. Wu and L. Wolfenstein, \PRL 73 1762 1994 ;
A. Rasin and J. P. Silva, \PRD 49 R20 1994 ;
K. Y. Lee and J. K. Kim, \MPL 10 1761 1995 ;
G. Cvetic, S. S. Hwang and C. S. Kim, \IJMP 14 769 1999 ;
G.~Cvetic, C.~S.~Kim and S.~S.~Hwang,
\PRD 58 116003 1998 .

\bibitem{collider} W. Khater and P. Osland, \NPB 661 209 2003 ;
Y.-B. Dai, C.-S. Huang, J.-T. Li, and W.-J. Li, \PRD 67 096007 2003 .

\bibitem{flavour} J.-F. Cheng, C.-S. Huang, and X.-h. Wu, \PLB 585 287 2004 ;
Y.-B. Dai, C.-S. Huang, J.-T. Li, W.-J. Li, \PRD 67 096007 2003 ;
G. Erkol, and G. Turan, \NPB 635 286 2002 ;
Y.-L. Wu and L. Wolfenstein, \PRL 73 2809 1994 .

\bibitem{GW} S. L. Glashow and S. Weinberg, \PRD 15 1958 1977 .

\bibitem{gunion} J. F. Gunion, B. Grzadkowski, H.E. Haber, and J. Kalinowski,
\PRL 79 982 1997 ;
B. Grzadkowski, J. F. Gunion, and J. Kalinowski, \PRD 60 075011 1999 ;
B. Grzadkowski, J. F. Gunion, and J. Kalinowski, \PLB 480 287 2000 .

\bibitem{ginzburg} I. F. Ginzburg and I. P. Ivanov, \PRD 72 115010 2005 ;
I. F. Ginzburg and M. V. Vychugin,
Published in the proceedings of 16th International Workshop
on High Energy Physics and Quantum Field Theory (QFTHEP 2001),
Moscow, Russia, 6-12 Sep 2001,
[arXiv] hep-ph/0201117.

\bibitem{kim}  K. Y. Lee, C. S. Kim, C. Yu,
To appear in the proceedings of International Conference on
Linear Colliders (LCWS 04), Paris, France, 19-24 Apr 2004,
[arXiv] hep-ph/0409009.

\bibitem{pomarol} A. Mendez and A. Pomarol, \PLB 272 313 1991 .


\end{thebibliography}
\end{document}